\documentclass[a4paper,conference]{IEEEtran}
\IEEEoverridecommandlockouts

\usepackage{cite}
\usepackage{amsmath,amssymb,amsfonts}
\usepackage{algorithmic}
\usepackage{graphicx}
\usepackage{textcomp}
\usepackage{xcolor}
\usepackage{tabularx}
\usepackage{placeins}
\usepackage{caption}

\def\BibTeX{{\rm B\kern-.05em{\sc i\kern-.025em b}\kern-.08em
    T\kern-.1667em\lower.7ex\hbox{E}\kern-.125emX}}
\begin{document}

\title{Significant Wave Height Estimation Incorporating Second-Order Scattering\\
\thanks{Natural Sciences and Engineering Research Council (NSERC) grant, Discovery Grant number RGPIN-2020-07155, presented to Dr. Reza Shahidi.}
}

\author{\IEEEauthorblockN{1\textsuperscript{st} Senal Chandrasekara}
\IEEEauthorblockA{\textit{Electrical and Computer Engineering} \\
\textit{Memorial University of Newfoundland}\\
St. John's, NL, Canada \\
cchandraseka@mun.ca}
\and
\IEEEauthorblockN{2\textsuperscript{nd} Reza Shahidi}
\IEEEauthorblockA{\textit{Electrical and Computer Engineering} \\
\textit{Memorial University of Newfoundland}\\
St. John's, NL, Canada \\
rshahidi@mun.ca}
}

\maketitle


\begin{abstract}
Traditional significant wave height (SWH) estimation from HF radar
typically relies on spectral analysis of the received radar signals. This process was previously simplified by establishing a linear relationship between 
SWH and the standard deviation of received HF radar voltages under first-order scattering. 
Building on this approach, this paper presents a physics-informed regression model that incorporates 
second-order scattering effects through a quadratic formulation derived from a Neumann expansion. 
The proposed method is evaluated using HF radar data collected in July 2018 at Argentia, Newfoundland, 
with collocated buoy measurements as ground truth. The model achieves a minimum root-mean-square error (RMSE) 
of approximately 19 cm.
\end{abstract}



\section{Introduction}

In \cite{shahidi_gill_2018} and \cite{shahidi_gill_2020}, a non-standard change of variables was used to show that, under single-scattering conditions, 
the standard deviation of the voltages received by an HF radar system from a given ocean patch is approximately linearly proportional 
to the average SWH within that region. More generally, these studies established that the ocean surface elevation 
$\xi$ at a patch centered at position $(\rho, \theta)$ in polar coordinates is approximately linearly related to the electric voltage 
measured from that patch at time $t$:

\begin{equation}
	E_{0,n}^+(\rho, \theta, t) \propto \xi(\rho, \theta, t)
	\label{LinProp}
\end{equation}

The upper envelope of the ocean surface elevation is known to follow a Rayleigh distribution \cite{tayfun2008distributions}. Consequently, 
the upper envelope of the magnitudes of the received voltages is also expected to approximately follow a Rayleigh distribution. In this work, 
Equation~\ref{LinProp} together with the statistical properties of the order statistics of the Rayleigh distribution are used to show that ranked 
voltage samples can be directly employed to estimate SWH. This leads to a significantly simpler approach for determining 
SWH using HF radar voltage measurements.

Recent progress in SWH estimation has increasingly relied on machine learning techniques and hybrid modeling strategies. 
Support Vector Machines, which are effective for relatively small datasets and nonlinear decision boundaries, have demonstrated
strong performance for long lead-time forecasting. Artificial Neural Networks (ANNs), on the other hand, provide greater modeling flexibility, 
and combining the two approaches has been shown to produce robust hybrid predictive models \cite{BERBIC2017331}.

Long Short-Term Memory (LSTM) networks have also been successfully applied to capture the temporal dynamics of SWH. 
When combined with architectures such as Bidirectional Gated Recurrent Units (BiGRUs) and Convolutional Neural Networks (CNNs), 
LSTMs can improve prediction accuracy by simultaneously modeling temporal dependencies and extracting spatial 
features \cite{AHMED2024111003, ZHANG2024102312, Chandrasekara2022SVMWeather}. More broadly, hybrid ML frameworks often outperform individual 
models by leveraging complementary strengths in data representation and learning \cite{fmars_2022}, \cite{ABBAS2024103919}. 
Furthermore, incorporating environmental variables such as wind speed has been shown to enhance predictive capability, highlighting the importance 
of meteorological inputs in modeling wave dynamics \cite{BERBIC2017331}, \cite{ALTUNKAYNAK20041245}, \cite{Barbara_2008}.

Extreme Learning Machines (ELMs) have attracted attention due to their relatively low computational complexity. 
Comparative studies involving ELMs, ANNs, and Support Vector Regressors (SVRs) indicate that ELM-based approaches can 
outperform traditional methods in certain scenarios \cite{Shahaboddin_2020}. Hybrid techniques such as 
Wavelet-PSO-ELM (WPSO-ELM) further improve prediction performance by combining wavelet-based feature extraction with 
particle swarm optimization (PSO) \cite{KALOOP2020107777}.

Despite these developments, prediction performance tends to degrade for longer forecasting horizons, particularly for 
LSTM-based methods \cite{fmars_2023}, \cite{fmars_2022}. In addition, many models struggle 
to maintain reliability under extreme sea conditions. In contrast, Feed-forward Neural Networks (FNNs), which employ simpler 
network structures, have demonstrated greater robustness in such scenarios \cite{Barbara_2008}.


\section{Proposed Method}
Prior work established that, only considering first-order scattering, the received HF radar voltage magnitude from a given ocean patch is proportional to 
the magnitude of the surface elevation \cite{shahidi_gill_2018, shahidi_gill_2020, shahidi_chandrasekara_2025}. 
These studies showed that the magnitude of the ocean surface height $\xi$ at any patch, centered at position ($\rho$, $\theta$) in polar coordinates, is linearly proportional to the magnitude of the electric voltage values received from that patch at any given time $t$:

\begin{equation}
	|E_{0,n}^+(\rho, \theta, t)| \propto |\xi(\rho, \theta, t)|
	\label{LinPropMagnitude}
\end{equation}

The random variable $|\xi|$ follows a distribution that can be interpreted as a
``folded'' Rayleigh distribution, and since $\xi$ has zero mean and may be reasonably assumed to be symmetric about
its mean value. The lower envelope of $|\xi|$ follows the same distribution as
the upper envelope, except that the mean of the lower envelope is the negative
of the mean of the upper envelope. Consequently, taking the absolute value of
$\xi$ results in an upper envelope equal to the larger of two quantities: the
upper envelope of the original $\xi$ and the negated lower envelope of $\xi$.
Equivalently, the resulting upper envelope can be viewed as the maximum of two
Rayleigh-distributed random variables.

A Rayleigh-distributed random variable with scale parameter $\mu$ has the
following cumulative distribution function (CDF):
\begin{equation}
	F_{Rayleigh}(x; \mu) = 1-e^{-x^2 / (2 \mu^2)}
\end{equation}
More generally, if two random variables have cumulative distribution functions
$F_1$ and $F_2$, respectively, then the cumulative distribution of the random
variable defined as the maximum of the two is given by
\begin{equation}
	F_{max} = F_1 \cdot F_2
\end{equation}
Applying this result to the envelopes described above, the CDF of the upper
envelope of $|\xi|$ becomes
\begin{equation}
	F_{\xi,upper}(x; \mu) = (1-e^{-x^2 / (2 \mu^2)})^2
	\label{CDF_upper}
\end{equation}
Since, for a single-scatter case, the magnitude of the received electric voltage
is linearly proportional to $|\xi|$ as expressed in
Equation~\ref{LinPropMagnitude}, it follows that for some positive real
constant $K$,
\begin{equation}
	F_{E,upper}(x; \mu) = (1-e^{-K x^2 / (2 \mu^2)})^2
	\label{CDF_upperE}
\end{equation}

\begin{eqnarray}
	\mathbb{E}\left[E^{+}_{0,n,(upper,k,N)}\right] = \mu \sqrt{\frac{2\ln\left(\frac{1}{1-\sqrt{\frac{k}{N}}}\right)}{K}} \label{CDF_inv}.
\end{eqnarray}

\noindent Since $H_s$ is known to be directly proportional to $\mu$, this implies that:

\begin{equation}
	E_{0,n}^{+} \propto \mu \propto H_s,
	\label{muprop}
\end{equation}

\noindent which motivates the estimation of $H_s$ using linear combinations of the
high-ranked voltage magnitudes, up to the first-order scatters. 

However, in practical HF radar measurements,
second-order scattering components introduce nonlinear distortions into the
received voltages, thereby limiting the accuracy of purely linear estimators.
When these effects are taken into account, the received complex voltage may be
modeled up to second order using a Neumann expansion \cite{shahidi_hashemi_2023},
leading to

\begin{equation}
	E_{0,n}^{+} \simeq AH_s^{2} + BH_s + C
	\label{secondOrderQuadratic}
\end{equation}


The coefficients $A$, $B$, and $C$ can be determined through a
calibration step using training data consisting of paired
HF radar voltage samples and collocated buoy measurements
of SWH. Specifically, a quadratic
least-squares regression can be performed between the radar
voltages and the corresponding buoy-measured wave heights
to estimate the coefficients of the polynomial model:

\begin{equation}
	AH_s^{2} + BH_s + (C - E_{0,n}^{+}) \simeq 0
	\label{secondOrderSubjected}
\end{equation}

\noindent Solving this yields

\begin{equation}
	H_s = \frac{-B \pm \sqrt{B^2 - 4A(C - E_{0,n}^{+})}}{2A}.
	\label{solvingForHs}
\end{equation}
	
Had the solutions for $H_s$ resulted from solving this equation been complex in nature, the absolute value of the real component of the resulted complex root can be considered. 

\begin{equation}
	\Re(H_s) = \left| \frac{-B}{2A} \right|
	\label{realPart}
\end{equation}	

Since $H_s$ is inherently real-valued, complex-valued
solutions obtained from the quadratic inversion are not physically admissible as
direct estimates of wave height. Such imaginary components can be attributed to
measurement noise, imperfect calibration of the coefficients, or limitations of
the second-order approximation. Therefore, the real component of the recovered
root is taken as the physically meaningful estimate.

Applying this quadratic inversion across samples produces a corrected data grid intended to suppress second-order 
scatter effects and restore behavior consistent with the first-order regime assumed in \cite{shahidi_chandrasekara_2025}. The corrected voltage magnitudes within 
each time segment are then sorted, and the top $M$ ranked samples are retained as robust features. The estimator is trained by solving the least-squares problem:

\begin{equation}
	\mathbf{w_{opt}} = \mathbf{\arg \min}||\mathbf{E}\mathbf{w} - \mathbf{h}||_2,
	\label{MatrixEqn}
\end{equation}
with the closed-form solution
\begin{equation}
	\mathbf{w_{opt}} = (\mathbf{E}^T \mathbf{E})^{-1} \mathbf{E}^T \mathbf{h}.
	\label{LSQRSoln}
\end{equation}
Predictions are computed via

\begin{equation}
\mathbf{h_{opt}} = \mathbf{E} \mathbf{w_{opt}},
\label{HsEst}
\end{equation}

\noindent and the non-negativity of SWH is enforced using the ReLU function. 


\section{Methodology}

The raw voltages recorded by the HF radar antennas include second-order
scattering contributions that exhibit quadratic behavior. These effects
can be addressed by solving Equation~\ref{secondOrderSubjected}. Each sample in the training dataset is used to estimate a quadratic
relationship between the radar voltages and the corresponding buoy-
measured SWHs using a least-squares fit. The
resulting coefficients $A$, $B$, and $C$ are then used to recover
estimates of $H_s$ according to Equation~\ref{solvingForHs}. When
complex-valued solutions arise, the magnitude of the real component,
computed using Equation~\ref{realPart}, is retained. The recovered
values form an empirical data grid of wave-height estimates that
partially compensates for second-order scattering effects.

As established previously, the expected value of any ordered statistic
of the received voltage magnitudes is linearly proportional to the
SWH $H_s$. The ordered statistics of the empirical
data grid are therefore assumed to be arranged in ascending order so
that

\begin{eqnarray}
	& \left[E^{+}_{0,n,(upper,1,N)}\right]  \leq 
	\left[E^{+}_{0,n,(upper,2,N)}\right] \leq\nonumber\\
	\ldots & \leq
	\left[E^{+}_{0,n,(upper,N-1,N)}\right] \leq
	\left[E^{+}_{0,n,(upper,N,N)}\right] 
	\label{SortedVoltages}
\end{eqnarray}

Order statistics with higher indices correspond to larger voltage
magnitudes and are therefore expected to be less sensitive to noise.
Consequently, a more reliable estimate of $\mu$, and hence the
SWH, can be obtained by forming a linear combination
of the ordered statistics in which higher-ranked samples receive greater
weight, rather than relying on a single statistic for each hour.
Furthermore, instead of using all derived voltage magnitudes, only a
subset is selected in order to reduce computational complexity when
determining the coefficients of the linear combination. Specifically,
the $M$ highest-ranked empirical voltage magnitudes are retained, where
$1 \leq M \leq N$ and $N$ represents the total number of time samples
within the given interval.

Suppose that the dataset contains $S$ segments of HF radar measurements,
each associated with a corresponding ground-truth value obtained from a
wave buoy or from a calculated estimate (e.g., Barrick's formula
\cite{barrick1977extraction}). Determining the optimal weights then
reduces to solving the following matrix equation in the least-squares
sense:

\begin{equation}
	\mathbf{w_{opt}} = \mathbf{\arg \min}||\mathbf{E}\mathbf{w} - \mathbf{h}||_2,
	\label{MatrixEqn}
\end{equation}

where the entries of the matrix $\mathbf{E}$ are defined by

\begin{equation*}
\mathbf{E}_{ij} = E^+_{0,n,(upper,N+1-j,N)}(t_i),
\end{equation*}

with $N+1-j$ representing the rank of the voltage magnitude within the
time interval beginning at $t_i$. To compute the optimal weight vector
$\mathbf{w_{opt}}$ in Equation \ref{MatrixEqn}, $K$ data segments are
selected such that $K \geq M$ and $1 \leq i \leq K$, while the $M$
largest ranked voltage magnitudes
$E^+_{0,n,(upper,N+1-j,N)}$ with $1 \leq j \leq M$ are retained.
The vector $\mathbf{w}$ contains the weights $\mathbf{w}_j$ used to form
a linear combination of the $M$ highest-ranked voltage magnitudes within
each data segment. The vector $\mathbf{h}$ contains the measured or
computed SWHs, where $h_i$ denotes the value for the
$i$th segment.

Equation \ref{MatrixEqn} admits a closed-form solution using the
pseudo-inverse, provided that $\mathbf{E}^T\mathbf{E}$ is invertible.
The resulting expression is

\begin{equation}
\mathbf{w_{opt}} = (\mathbf{E}^T \mathbf{E})^{-1} \mathbf{E}^T \mathbf{h}
\label{LSQRSoln}
\end{equation}

Equation \ref{LSQRSoln} is used to determine the optimal weight vector
$\mathbf{w_{opt}}$. These weights are then applied to estimate
SWHs by forming a linear combination of the $M$
largest ranked voltage magnitudes within each time interval. The sorted
voltage magnitudes are arranged into the rectangular matrix
$\mathbf{E}$, and the predicted wave heights $\mathbf{h_{opt}}$ are
computed through a simple matrix multiplication:

\begin{equation}
\mathbf{h_{opt}} = \mathbf{E} \mathbf{w_{opt}}
\label{HsEst}
\end{equation}

Higher-order statistics corresponding to the largest ranked magnitude
values closely approximate the upper envelope of the received voltage
signal. As a result, it is possible to bypass explicit upper-envelope
extraction and instead operate directly on the raw voltage magnitudes.

Finally, since the SWH must be non-negative, the
predicted values are passed through the ReLU activation function:

\begin{equation}
\mathbf{H_{s,est}} = \mathrm{ReLU}\mathbf{(h_{opt})}
\label{HsEst_ReLU}
\end{equation}


\section{Experiments}
The proposed method is evaluated by first solving for $H_s$ to obtain an
empirical grid of voltage magnitudes. For each time segment, these voltage
magnitudes are sorted in descending order. From this ordered set, the top
$M$ ranked voltage magnitudes are selected, where $M$ is varied from $1$ to
$N$, with $N$ denoting the total number of samples in a data segment. The
selected $M$ highest-ranked samples form the first row of the matrix
$\mathbf{E}$.

This procedure is repeated for the first $K$ data segments of the complete
HF radar dataset. To avoid an underdetermined system when solving for the
model weights, the constraint $K \geq M$ is enforced. Additionally, $K$
cannot exceed $S$, where $S$ represents the total number of available data
segments in the HF radar dataset.

For each valid pair of $M$ and $K$, the model is trained by computing the
optimal weight vector $\mathbf{w_{opt}}$ using Equation~\ref{LSQRSoln}.
These weights are then applied to estimate the SWH for
the remaining $N-M$ data segments.
The resulting predicted $H_s$ values are subsequently smoothed using a
sliding window of length $W$, where $W$ is chosen as an odd integer between
1 and 49. Only backward averaging is used so that the smoothing operation
can be performed in real time without requiring future measurements. This
backward averaging step may also be interpreted as a hidden layer in the
learning framework, with weights equal to the reciprocal of the averaging
window length and multiple high-ranked samples from consecutive segments
serving as inputs.

Finally, the RMSE is computed between the smoothed
predicted $H_s$ values and the corresponding ground-truth values obtained
from wave buoy measurements or from Barrick's formula for the final
$N-M$ HF radar data segments. The pair of parameters $M$ and $K$ that
produces the minimum RMSE is selected, and the associated weight vector
$\mathbf{w_{opt}}$ is taken as the optimal set of model parameters.

\begin{figure}[]
    \centering
    \includegraphics[height=3.5cm]{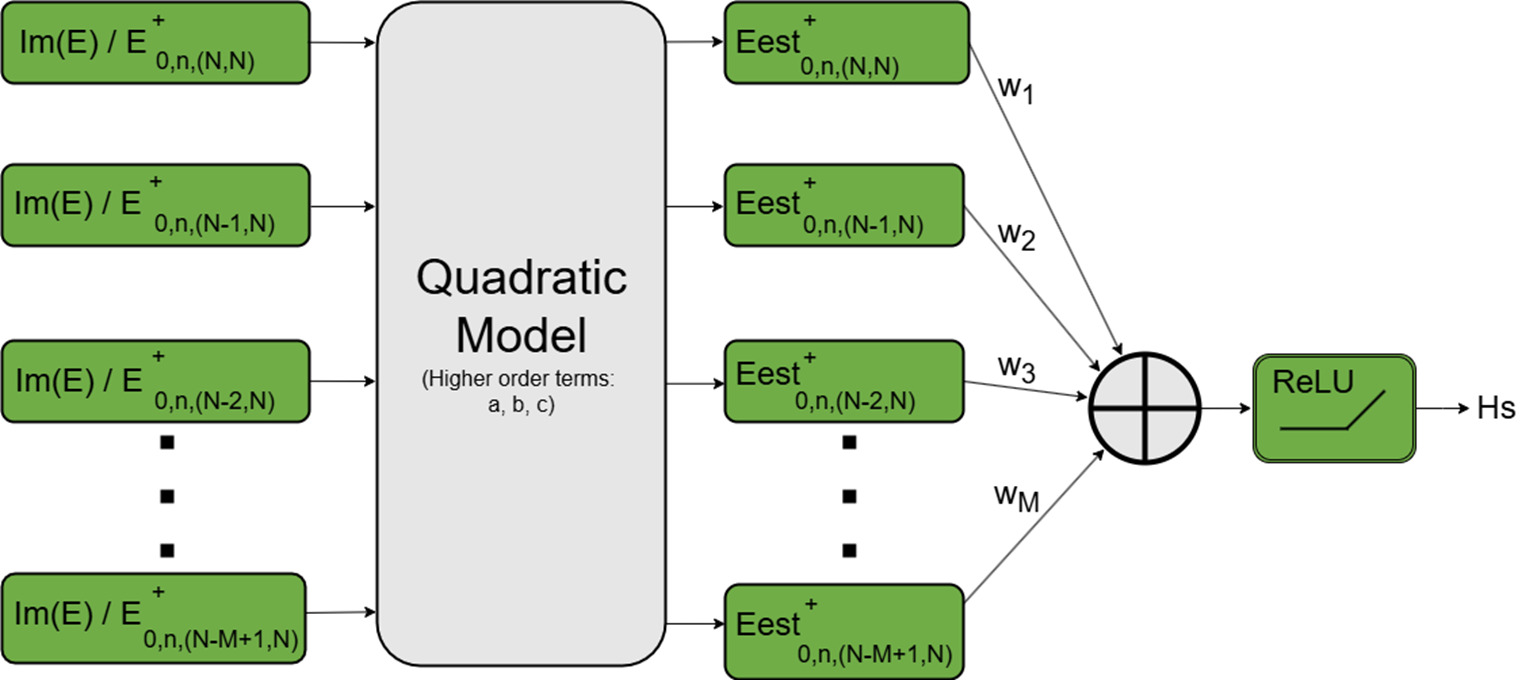}
    \caption{The quadratic inversion model architecture.}
    \label{QuadraticModel}
\end{figure}

\begin{figure}[]
    \centering
    \includegraphics[height=5.0cm]{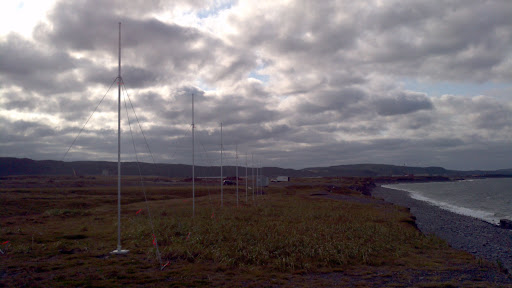}
    \caption{The HF radar site located in Argentia, NL, Canada.}
    \label{radar_site}
\end{figure}

\begin{figure}[]
    \centering
    \includegraphics[height=4.5cm]{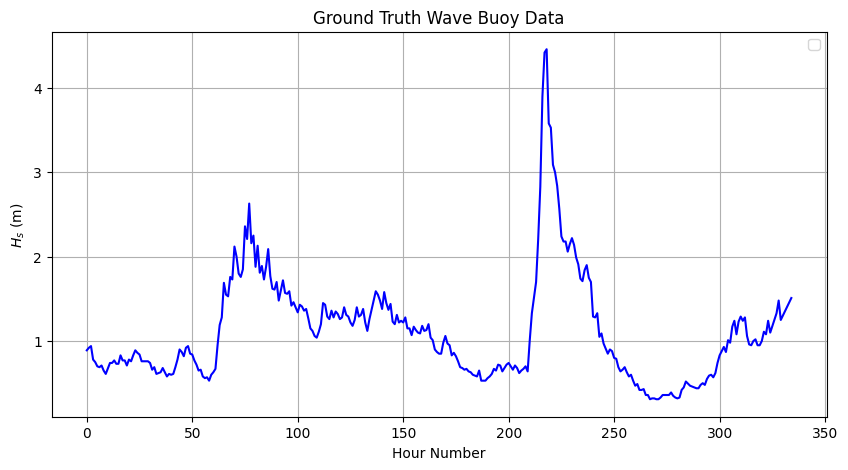}
    \caption{Groundtruth (buoy) data used to train model.}
    \label{buoy_data}
\end{figure}

These experiments were conducted using data collected from the Memorial
University HF radar site located in Argentia, Newfoundland. Fig.\ref{radar_site}
illustrates the 12-element HF radar receiver array which, after
beamforming, produced a dataset consisting of 4096 samples per hour
over a total duration of 335 hours. Fig.~\ref{buoy_data} shows the location of the ocean buoy (red star) relative to the radar
array (circled 'x' filled in green), which was used to obtain the ground-truth SWH
measurements for the same 335 hours. Fig.~\ref{buoy_data} presents the complete ground-truth dataset used in
this study.

\begin{figure}[]
    \centering
    \includegraphics[height=6.5cm]{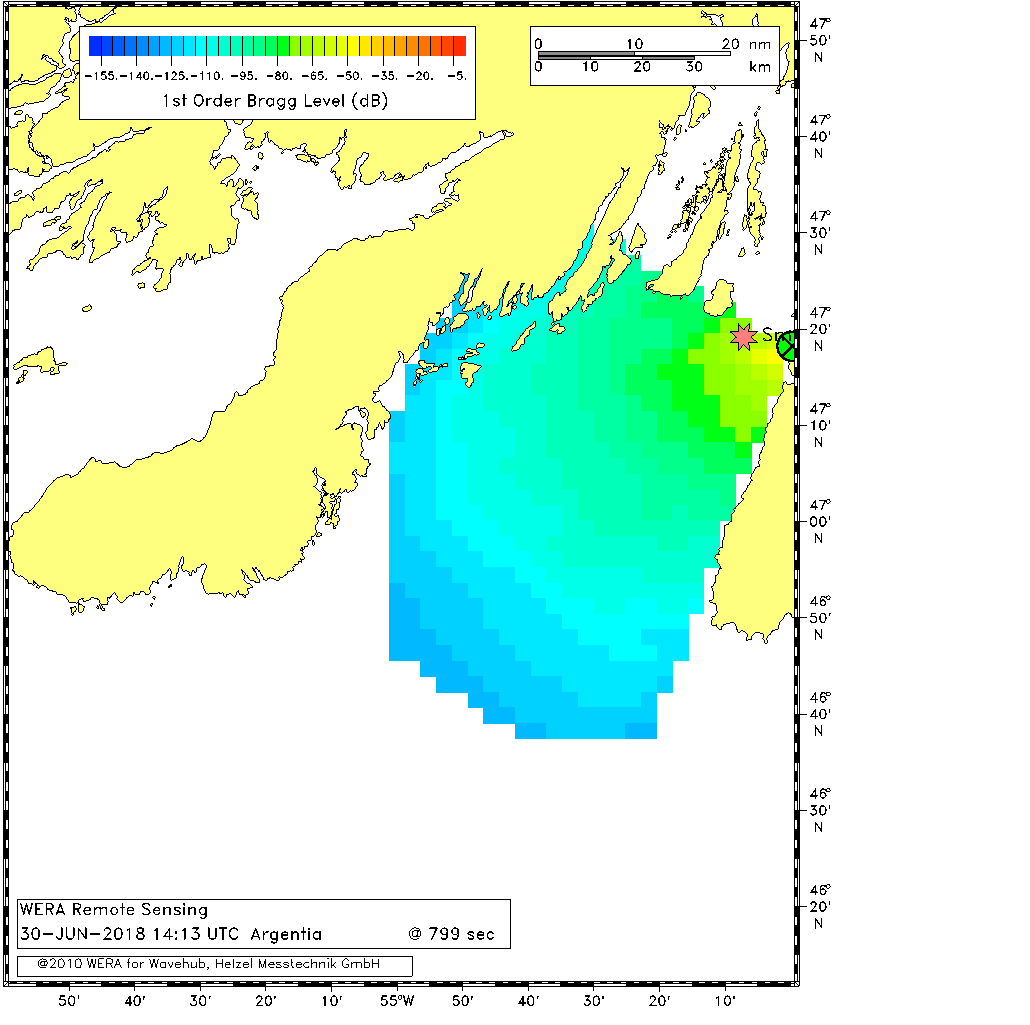}
    \caption{Location of the buoy (red star).}
    \label{buoy_site}
\end{figure}

\section{Results}

\begin{figure}[]
\centering
\includegraphics[height=4.8cm]{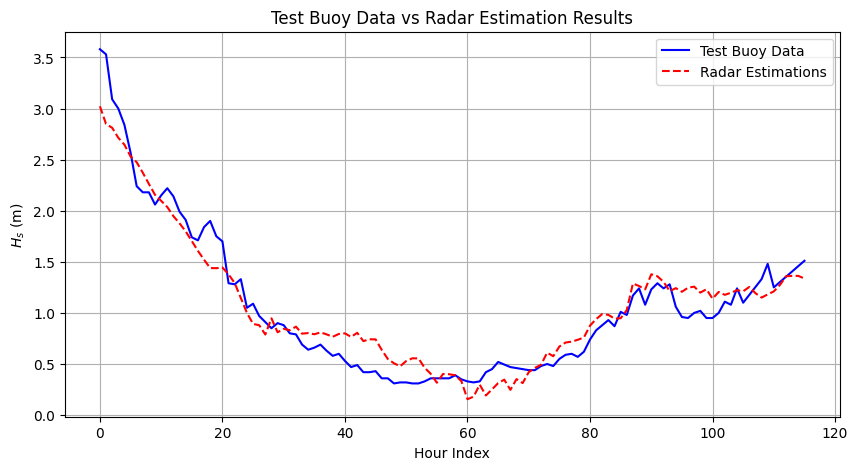}
\caption{Buoy measurements vs. predicted $H_s$ values.}
\label{HsPredictionsQuadratic}
\end{figure}

\begin{figure}[]
\centering
\includegraphics[height=8.0cm]{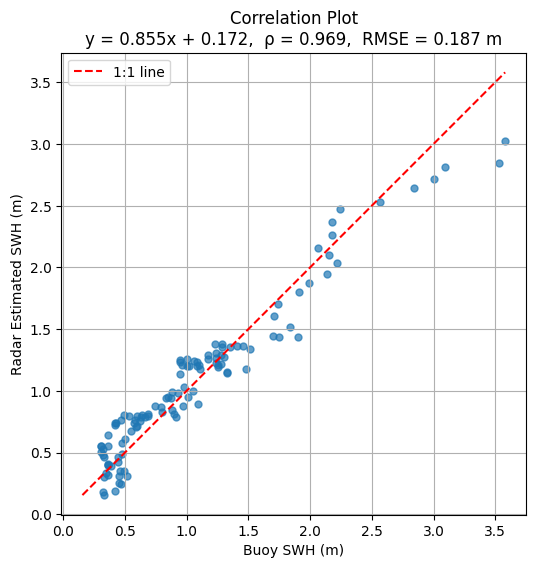}
\caption{Correlation between the predicted SWHs and the corresponding buoy measurements for the optimal parameter configuration.}
\label{cor_plot}
\end{figure}

Using magnitude, real, and imaginary components of the radar IQ samples, the model was trained to
achieve minimum testset RMSEs below 20 cm, with the imaginary component providing the best performance as observed in Table \ref{tab:results}. 
Fig.\ref{HsPredictionsQuadratic}, illustrating the performance of this approach on real-world radar voltage samples against ground-truth buoy data, shows strong agreement 
between predicted and buoy-measured $H_s$ on the test data, hence demonstrating consistent performance across the evaluated sea states. 
The correlation plot corresponding to the optimal parameter set with a correlation coefficient of ~0.97 set is demonstrated in Fig.~\ref{cor_plot}.
{
\captionof{table}{Performance Summary}
\label{tab:results}
\renewcommand{\arraystretch}{1.1}
\setlength{\tabcolsep}{4pt}
\begin{tabularx}{\columnwidth}{X c c}
\hline
\textbf{Input Data} &
\textbf{Rank Subset} &
\textbf{Min. RMSE (cm)} \\
\hline
Magnitudes & 219 & 19.4 \\
Imaginary components & 196 & 18.7 \\
Real components & 243 & 19.3 \\
\hline
\end{tabularx}
}

\section{Conclusion}
This paper presents an approach for estimating SWH from HF radar receiver
electric-field voltage time series by accounting for scond-order scatter effects. The method employs a quadratic mapping between radar voltages and buoy-measured wave heights, 
followed by root-based reconstruction, rank ordering of quadratic estimates, and linear least-squares combination with temporal smoothing. 
Experimental evaluation demonstrates that the proposed approach achieves close agreement with buoy observations, 
with minimum RMSEs in the range of 18 to 19 cm, and the lowest error of 18.7 cm obtained when using the imaginary components 
from the recevied HF radar IQ samples. These results confirm the effectiveness of the proposed framework for practical HF radar-based wave height estimation when 
second-order scattering effects are implicitly captured.

\section*{Acknowledgment}
The authors would like to acknowledge the support of the
Natural Sciences and Engineering Research Council (NSERC)
through Discovery Grant number RGPIN-2020-07155 to Dr.
Reza Shahidi.

\bibliographystyle{IEEEtran}
\bibliography{references}
\end{document}